\documentclass[aps,
prd,
twocolumn
]{revtex4-2}

\usepackage{graphicx}
\usepackage{dcolumn}
\usepackage{bm}
\usepackage{comment}
\usepackage{xcolor}
\usepackage{amsfonts}
\usepackage{amsmath}
\usepackage{slashed}

\begin{document}

\preprint{APS/123-QED}

\title{Simulating Majorana fermions in black hole with Ising Models}

\author{John Vienn A. Estremadura}
\email{jaestremadura@up.edu.ph}
 \affiliation{%
 National Institute of Physics, University of the Philippines Diliman, Philippines.
 }%
\author{Kristian Hauser A.~Villegas}%
 \email{kavillegas1@up.edu.ph}
\affiliation{%
 National Institute of Physics, University of the Philippines Diliman, Philippines.\\
}%

\date{\today}

\begin{abstract}
Quantum field theory (QFT) in curved spacetime has led to profound predictions, including the Unruh effect and Hawking radiation, yet their direct observation remains extraordinarily challenging because of their extremely weak signatures. Here, we show that the transverse-field Ising model provides a quantum simulator for Majorana fermions in a Schwarzschild black hole background. Remarkably, different coordinate representations of the same spacetime—Schwarzschild, tortoise, Kruskal, and conformally flat—map onto distinct microscopic Ising spin models. Despite their microscopic differences, these models converge in the continuum limit to the same Majorana field theory, exhibiting an emergent form of general covariance. This provides a rare example of a fundamental symmetry of general relativity arising as an emergent property of a condensed matter system. We further demonstrate how black hole particle production can be simulated and detected through spin correlation measurements, and discuss experimental platforms capable of realizing these models. Our work establishes a practical route for investigating fermionic QFT in curved spacetime using controllable quantum many-body systems and tabletop experiments.
\end{abstract}

\maketitle

\section{Introduction}\label{sec:intro}Quantum field theory (QFT) in curved spacetime provides a semi-classical framework that extends conventional QFT from flat Minkowski spacetime to curved backgrounds. Unlike a complete theory of quantum gravity, the spacetime geometry is treated as a classical field governed by general relativity, while matter fields are quantized. This approximation offers a powerful setting in which to investigate the interplay between quantum physics and gravitation, serving as an important bridge between two of the most successful yet fundamentally distinct theories in modern physics.

The study of QFT in curved spacetime has led to several profound insights into the nature of quantum fields in gravitational environments. It predicts phenomena such as the Unruh effect \cite{Unruh1976}, in which accelerated observers perceive the vacuum as a thermal state, and particle production in expanding universes \cite{Parker1968}, where the dynamical geometry of spacetime can generate particles from quantum fluctuations. Most notably, it provides the theoretical foundation for Hawking radiation, revealing that black holes emit thermal radiation and can gradually evaporate \cite{Hawking1975}. These discoveries have fundamentally altered our understanding of concepts such as particles, vacuum states, and horizons, demonstrating that they can depend on the geometry of spacetime and the observer's state of motion.

Beyond its foundational successes, QFT in curved spacetime plays a central role in modern cosmology and black-hole physics. Quantum fluctuations of fields in the early universe are believed to have been amplified by cosmic expansion, ultimately seeding the large-scale structure observed today. At the same time, the thermal properties of black holes uncovered within this framework have given rise to deep questions concerning entropy, information, and the quantum nature of gravity. As such, QFT in curved spacetime not only provides a mathematically rich framework for studying quantum effects in gravitational settings but also offers crucial clues toward a more fundamental theory unifying quantum mechanics and gravitation.

Despite its profound theoretical successes, the experimental observation of quantum field-theoretic effects in curved spacetime remains exceptionally challenging. Phenomena such as Hawking radiation and cosmological particle production typically arise only under extreme gravitational or cosmological conditions, and even then their signatures are exceedingly weak, placing them far beyond the reach of current laboratory experiments. Fortunately, it is now widely recognized that QFT emerges universally as an effective low-energy description of a broad class of quantum many-body systems. This insight has opened a promising avenue for the analog simulation of QFT in curved spacetime \cite{Barcelo2011AnalogueGravity, Almeida2023AnalogueGravity, Liu2025} using controllable condensed-matter and quantum-engineered platforms, including spin systems \cite{Horner2023, Forbes2023, Kinoshita2025, Kinoshita2025b, Calliari2025}, polaritonic systems \cite{Falque2025}, Bose–Einstein condensates \cite{Viermann2022}, ultracold atoms \cite{Cirac2010}, and a variety of other lattice \cite{Deger2023} and quantum many-body systems \cite{Yang2020}.

In these analog platforms, the long-wavelength or continuum limit gives rise to effective field theories whose dynamics can be mapped onto those of quantum fields propagating in curved spacetime backgrounds. By engineering suitable interactions, geometries, or time-dependent parameters, one can emulate aspects of gravitational physics in highly controllable laboratory settings. Such analog simulations provide a unique opportunity to investigate fundamental predictions of QFT in curved spacetime through table-top experiments, offering access to phenomena that would otherwise be inaccessible to direct experimental observation. Beyond their experimental utility, these systems also provide valuable insight into the emergence of spacetime concepts from underlying microscopic quantum degrees of freedom, thereby deepening our understanding of the relationship between quantum many-body physics, quantum fields, and gravity.

Recently, it was shown that suitably engineered spin systems can serve as analog simulators of quantum field theories of Majorana fermions in curved spacetime backgrounds \cite{Kinoshita2025}. In particular, the authors demonstrated how the Unruh effect can emerge in a spin model corresponding to an expanding universe. Building on this framework, we study the various tranverse-field Ising models that simulates the Majorana QFT in a black hole background. 

A particularly striking feature of our construction is that different coordinate representations of the same Schwarzschild black hole background generally give rise to distinct microscopic spin Hamiltonians. Despite these differences at the lattice level, we show that all such models are equivalent up to coordinate transformations in the continuum limit. Consequently, the continuum description exhibits an emergent general covariance that lies at the heart of general relativity. This result is especially noteworthy because the symmetry is not imposed at the microscopic level but instead emerges dynamically in the low-energy theory. While emergent Lorentz and gauge symmetries are familiar phenomena in condensed-matter and quantum many-body systems, examples of emergent general covariance are considerably rarer. Our work therefore provides a concrete realization of how spacetime coordinate invariance can arise from fundamentally non-relativistic microscopic degrees of freedom, offering new insights into the emergence of gravitational concepts in quantum many-body systems.

The structure of the paper is as follows. In Section \ref{sec:mappingreview}, we review the mapping of Majorana fermions in curved spacetime to microscopic spin systems, following the framework developed by Kinoshita et al. \cite{Kinoshita2025}. In Section \ref{sec:blackholebackground}, we construct families of spin models corresponding to the QFT of Majorana fermions in the background of a Schwarzschild black hole, using various coordinate systems—including Schwarzschild, tortoise, and Kruskal coordinates. Leveraging the freedom to choose coordinates and the emergence of general covariance, which we prove in Section \ref{sec:diffeomorphism}, we identify particularly convenient spin models where spatial dependence appears only in the transverse-field term. This simplification arises naturally in conformal coordinates such as the tortoise and Kruskal coordinates. We then employ these simplified models to propose how particle production in a black hole can be simulated and detected through measurements of various spin correlations. Finally, Section VI presents our conclusions and outlook, where we discuss potential experimental platforms for simulating Majorana QFT in a black hole background.

%
\section{Mapping Majorana in curved spacetimes into spin systems}
\label{sec:mappingreview}
We start by reviewing the results of Kinoshita et al.\cite{Kinoshita2025} on how to simulate Majorana fields in curved spacetimes using spin models. We will also follow their conventions and notations.

Let us consider the two-dimensional curved spacetime background whose metric can be written in the Arnowitt-Deser-Misner (ADM) formalism \cite{Poisson2004} as
\begin{align}
\label{eq:admmetric}
    ds^2=-\alpha(t,x)^2 dt^2 +\gamma(t,x)^2 [dx-\beta (t,x) dt]^2,
\end{align}
where $\alpha$, $\beta$, and $\gamma$ are called the metric functions. 

In two dimensions, the spin connection for the spinor field vanishes and the Lagrangian density is given by
\begin{align}
\label{eq:lagrangian1}
    \mathcal{L}=-i\sqrt{-g}\psi^T\gamma^0(\slashed{\partial}-m)\psi,
\end{align}
where $\psi=(\psi_1,\psi_2)^T$ with $\psi_j$ being Grassmann-valued variables and satisfying $\psi_j^\dagger=\psi_j$. That is, the components $\psi_i$ obey the Majorana condition.

Introducing the complex variables
\begin{align}
    \chi=\psi_2-i\psi_1,\;\;\chi^\dagger=\psi_2+i\psi_1
\end{align}
and redefining the field
\begin{align}
    \chi=\gamma^{-1/2}e^{i\zeta/2}\Psi,
\end{align}
we can obtain the corresponding Hamiltonian density from the Lagrangian Eq. \eqref{eq:lagrangian1} for a Majorana fermion field $\Psi$ of mass $m$:
\begin{align}
\label{eq:QFT}
    \mathcal{H}=&-\frac{\alpha}{2\gamma} \cos{\zeta} \left(\Psi^{\dagger} \partial_x \Psi^{\dagger}-\Psi \partial_x \Psi\right)\nonumber\\
    &+i\frac{\alpha}{2\gamma} \sin{\zeta} \big(\Psi^{\dagger} \partial_x \Psi^{\dagger}+\Psi \partial_x \Psi\big)\nonumber\\
    &-i\frac{\beta}{2} \left(\Psi^{\dagger} \partial_x \Psi+\Psi \partial_x \Psi^{\dagger}\right)\nonumber\\
   & -\left[m\alpha-\frac{1}{2}\left(\partial_t \zeta+\beta\partial_x \zeta\right)\right]\Psi^{\dagger} \Psi.
\end{align}
In writing this equation, we omit the spacetime arguments $(t,x)$ of the functions $\alpha$, $\beta$, $\gamma$, and $\zeta$ for notational brevity. The function $\zeta(t,x)$ represents the phase of the complex fermion field $\Psi$.

We want to match Eq. \eqref{eq:QFT} to a continuum limit of the microscopic spin model described by the Hamiltonian of the form
\begin{align}
\label{eq:hamiltonianspin}
    H=-\sum_j\left[\sum_{a,b=\pm}J^{ab}_j (t)\sigma^a_j\sigma^b_{j+1}+h_j (t)\sigma^z_j\right],
\end{align}
where $\sigma^\pm_j\equiv(\sigma^x_j\pm i\sigma^y_j)/2$ and $\sigma^{x,y,z}_j$ are the Pauli spin matrices, $j$ labels the lattice site, $J^{ab}_j$ is the exchange interaction, and $h_j$ is the transverse magnetic field.

Equation \eqref{eq:hamiltonianspin} can be written in terms of fermion operators $c_j$ and $c_j^\dagger$ via the Jordan-Wigner transformation
\begin{align}
\label{eq:jordanwigner}
    \sigma^+_j=&\prod_{l=1}^{j-1}(1-2c^\dagger_lc_l)c_j\\
    \sigma^-_j=&\prod_{l=1}^{j-1}(1-2c^\dagger_lc_l)c_j^\dagger\\
    \sigma^z_j=&1-2c^\dagger_jc_j.
\end{align}

Here we follow the Jordan-Wigner transformation used in \cite{Kinoshita2025}, which is the opposite of the usual one where $\sigma^+_j\propto c_j^\dagger$ and $\sigma^-_j\propto c_j$. These two conventions are related by the particle-hole transformation.

We then take the continuum limit by defining the field
\begin{align}
    \Psi(x_j)\equiv\frac{c_j}{\sqrt{\epsilon}},
\end{align}
where $\epsilon$ is the lattice spacing, and writing the hopping terms as
\begin{align}
    c_j^\dagger c_{j+1}=&\epsilon\Psi^\dagger(x_j)\Psi(x_j+\epsilon)\nonumber\\
    \approx&\epsilon\Psi^\dagger(x_j)\Psi(x_j)+\epsilon^2\Psi^\dagger(x_j)\partial_x\Psi(x_j),
\end{align}
to rewrite the Hamiltonian Eq. \eqref{eq:hamiltonianspin} in terms of the low-energy excitations $\Psi$. The resulting effective low-energy Hamiltonian has the same form as the Hamiltonian in Eq. \eqref{eq:QFT}. By comparing the two, we can relate the metric functions $\alpha$, $\gamma$, and $\beta$, along with the Majorana phase $\zeta$, with the parameters of the spin Hamiltonian $J^{ab}_j$ and $h_j$ \cite{Kinoshita2025}. Writing the spin Hamiltonian in terms of the former set of parameters, we have
\begin{align}
\label{spinsys}
H=&-\frac{1}{4\epsilon}\sum_{j=1}^{N}
    \bigg\{\left(\frac{\alpha}{\gamma} \cos{\zeta}+p\right)\sigma_j^x \sigma_{j+1}^x
    \nonumber\\
    &-\left(\frac{\alpha}{\gamma} \cos{\zeta}-p\right)\sigma_j^y \sigma_{j+1}^y
    -\left(\beta+\frac{\alpha}{\gamma} \sin{\zeta}\right)\sigma_j^x \sigma_{j+1}^y
    \nonumber\\
    &+\left(\beta-\frac{\alpha}{\gamma} \sin{\zeta}\right)\sigma_j^y \sigma_{j+1}^x
    +\big[2p-\epsilon(\partial_x p+2m\alpha\nonumber\\
    &-\partial_t \zeta-\beta\partial_x \zeta)\big]\sigma_j^z\bigg\}. 
\end{align}
For different values of parameters, this describes a family of spin chain models that converge to Eq. \eqref{eq:QFT} in the continuum limit. Here, $p$ and $\zeta$ are free functions of $t$ and $x$, in the sense that different choices of these functions lead to the same Majorana field theory in the continuum limit. In the spin model, $\zeta(t,x)$ corresponds to the rotation of the spin model about the $z$-axis: $\sigma_j^{\pm}\to e^{i\zeta/2}\sigma_j^{\pm}$. Without loss of generality, we will choose $\zeta(t,x)=0$ throughout this paper, which simplifies the continuum and spin Hamiltonians to
\begin{align}
    \label{continuum_nozeta}
    \mathcal{H}=&-\frac{\alpha}{2\gamma} \left(\Psi^{\dagger} \partial_x \Psi^{\dagger}-\Psi \partial_x \Psi\right)
    -i\frac{\beta}{2} \left(\Psi^{\dagger} \partial_x \Psi\right.\nonumber\\
    &\left.+\Psi \partial_x \Psi^{\dagger}\right)-m\alpha\Psi^{\dagger} \Psi,\\
    \label{nozeta}
     H=&-\frac{1}{4\epsilon}\sum_{j=1}^{N}
    \bigg\{\left(\frac{\alpha}{\gamma}+p\right)\sigma_j^x \sigma_{j+1}^x
    -\left(\frac{\alpha}{\gamma}-p\right)\sigma_j^y \sigma_{j+1}^y\nonumber\\
    &-\beta\left(\sigma_j^x \sigma_{j+1}^y+\sigma_j^y \sigma_{j+1}^x\right)
    +\left[2p-\epsilon\left(\partial_x p+2m\alpha\right)\right]\sigma_j^z\bigg\}, 
\end{align}
respectively.

Hence, the target metric background can be realized through the functions $\alpha$, $\beta$, and $\gamma$ by suitably engineering the spin-exchange interactions and transverse magnetic field appearing in Eq.~\eqref{nozeta}. In this way, the microscopic parameters of the spin model directly determine the effective curved spacetime experienced by the emergent Majorana fermions.

In contrast to the phase $\zeta$, the function $p(t,x)$ cannot be eliminated through a redefinition of the spin operators. Consequently, different choices of $p(t,x)$ correspond to genuinely distinct microscopic spin Hamiltonians. Remarkably, despite these microscopic differences, all such models flow to the same Majorana field theory in the continuum limit. This freedom offers significant practical advantages. By appropriately choosing $p(t,x)$, one can tailor the microscopic model to simplify analytical calculations, reveal particular physical features more transparently, or facilitate experimental implementation while preserving the same low-energy continuum description. For QFTs with open boundaries (such are the ones considered in this paper), however, the value of the function $p(t,x)$ at the boundaries must follow a specific boundary condition to avoid fermion doubling \cite{kinoshita2026open}. In particular, for a QFT defined in the region $[0,l]$, the allowed boundary conditions in the continuum theory is reproduced by the spin model if the value of $p(t,x)$ at the boundaries satisfy:
\begin{align}
    p|_{x=0}=-s\sqrt{\frac{\alpha^2}{\gamma^2}-\beta^2},\quad
    p|_{x=l}=s'\sqrt{\frac{\alpha^2}{\gamma^2}-\beta^2},
\end{align}
where $s,s'=\pm 1$, which can be chosen freely at the boundary. Despite these constraints on $p$ at the boundary, its bulk values can be chosen freely, provided that it is sufficiently smooth near the boundaries \cite{kinoshita2026open}.

In the next section, we will use this mapping to study the various spin models that can simulate Majorana field in the black hole background.

\section{Spin systems simulating Majorana in blackhole background}
\label{sec:blackholebackground}
We have seen from the Hamiltonian in Eq. \eqref{nozeta} that we have a family of spin models converging to the same continuum limit. Furthermore, these models depend on the metric functions $\alpha$, $\beta$, and $\gamma$, whose forms depend on the particular coordinates used. In this section, we study and compare the different spin models generated using different coordinate systems of the Schwarzschild black hole.



\subsection{Spin models from Schwarzschild coordinates}
\label{sec:theory}
We start with the Schwarzschild coordinates $(t,x)$, where $x\geq 0$ is the usual radial distance. The metric is given by
\begin{equation}
\label{eq:schwarzschildmetric}
    ds^2=-f(x)dt^2+\frac{1}{f(x)} dx^2,
\end{equation}
where $f(x)=1-r_s/x$ and $r_s=2GM$ is the Schwarzschild radius with $M$ the black hole mass and $G$ the gravitational constant. Notice that for this toy model, we suppressed the angular coordinates in order to have a two-dimensional spacetime. 

Let us derive the spin model generated by Eq. \eqref{nozeta} that has Majorana QFT in this spacetime background in the continuum limit. To do this, we split the spacetime into two regions: outside the black hole $x>r_s$ and inside the black hole $x<r_s$. 

We first focus on the exterior region of the black hole, $x>r_s$, where $f(x)$ is positive. Comparing Eq.~\eqref{eq:schwarzschildmetric} with the general two-dimensional metric given in Eq.~\eqref{eq:admmetric}, we identify
\begin{equation}
\label{vielbeins}
    \alpha(t,x)^2=f(x),\quad
    \gamma(t,x)^2=\frac{1}{f(x)},\quad
    \beta(t,x)=0.
\end{equation}

Using these, we obtain the family of spin models
\begin{align}
H_{out}=&-\frac{1}{4\epsilon}\sum_{j=n}^{N}
    \bigg\{(p+f)\sigma_j^x \sigma_{j+1}^x
    +(p-f)\sigma_j^y \sigma_{j+1}^y\nonumber\\
    &+\left[2p-\epsilon(\partial_x p+2m\sqrt{f})\right]\sigma_j^z\bigg\} ,\label{blackshield}
\end{align}
where we define $n$ to be the location of the horizon $r_s=\epsilon n$.

\begin{figure*}[t]
    \centering

    \includegraphics[trim=0cm 6.6cm 0cm 2cm, clip, width=1.0\textwidth]{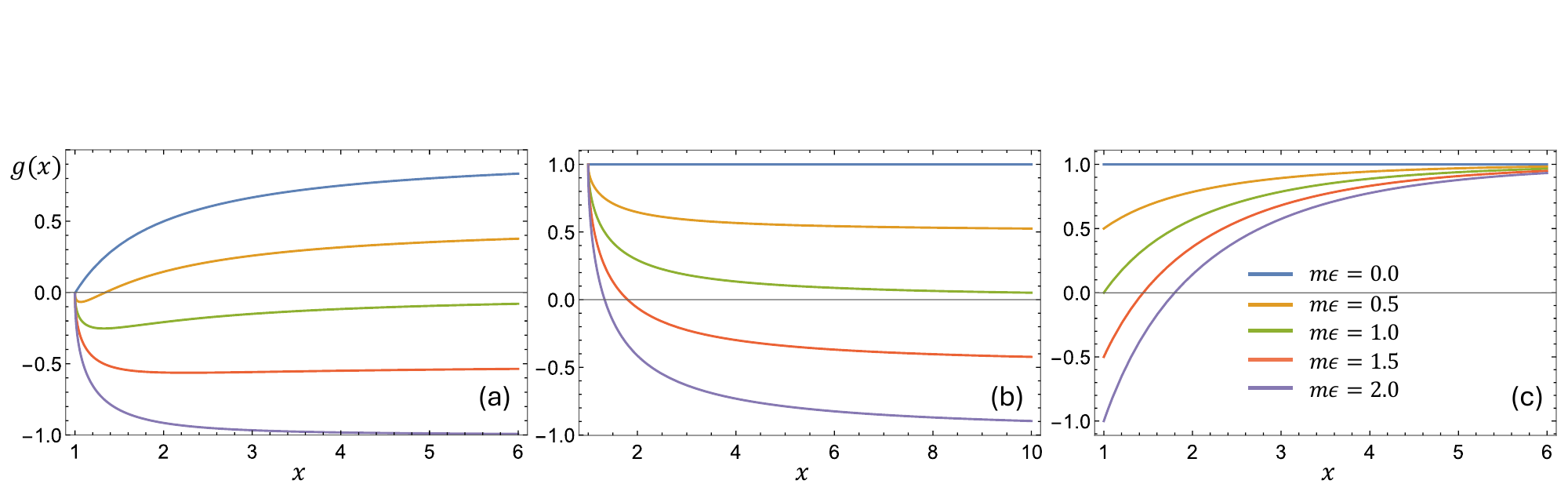}

    \caption{Transverse magnetic field as a function of position for spin systems generated using (a) Schwarzschild, (b) tortoise, and (c) Kruskal coordinates, plotted for various values of the parameter $m\epsilon$. The black hole horizon is set at $r_s=1$.}
    \label{fig:P1_ret_array}
\end{figure*}

We now take advantage of the fact that $p(t, x)$ is a free function—meaning that the continuum-limit quantum field theory is independent of its specific form—to select a particular spin model from this family that is simpler and analytically tractable. Specifically, we choose $p(t, x) = \alpha / \gamma = f$, which leads to the following spin Hamiltonian:
\begin{equation}
\label{eq:hamiltonianout}
H_{out} =-\frac{1}{2\epsilon} \sum_{j=n}^{N}\left\{f_j\sigma_j^x \sigma_{j+1}^x+g_j \sigma_j^z\right\},
\end{equation}
where $f_j=f(x_j)$ is the discretized version of the function $f(x)$, $x_j=\epsilon j$ for $j=n,n+1,\cdot\cdot\cdot N$ is the site position, and
\begin{align}
\label{eq:transversefield1}
    g_j=f_j-\epsilon r_s/(2x_j^2)-m\sqrt{f_j}
\end{align}
is the transverse magnetic field acting on the spin system. The spatial profile of the transverse magnetic field is shown in Fig. \ref{fig:P1_ret_array}(a). As evident from the figure, the transverse field vanishes at the horizon and approaches a constant value asymptotically at large distances. Interestingly, for a fixed lattice spacing $\epsilon$, increasing the Majorana mass $m$ leads to a reversal in the direction of the transverse field.

Inside the black hole, the roles of $x$ and $t$ in the Schwarzschild coordinates are switched. Here $f(x)<0$ so that in Eq. \eqref{eq:schwarzschildmetric}, $x$ becomes timelike while $t$ becomes spacelike. A comparison of Eqs. \eqref{eq:admmetric} and \eqref{eq:schwarzschildmetric} now yields
\begin{equation}
\label{eq:vielbeins2}
    \alpha(t,x)^2=-\frac{1}{f(x)},\;\;
    \gamma(t,x)^2=-f(x),\;\;
    \beta(t,x)=0.
\end{equation}
Comparing these expressions with Eq.~\eqref{vielbeins}, we find that $\alpha$ and $\gamma$ are exchanged, together with the sign change of $f(x)$. This is a direct consequence of the interchange of the coordinates $x$ and $t$ across the horizon, and hence of the corresponding exchange in the roles of $\alpha$ and $\gamma$.

The roles of the functions $\alpha$ and $\gamma$ in the spin Hamiltonian Eq. \eqref{nozeta} must also be switched before using Eq. \eqref{eq:vielbeins2}. Overall, these manipulations have the effect of effectively changing $f_j\rightarrow -f_j$ in Eq. \eqref{eq:hamiltonianout} and we get
\begin{align}
\label{eq:hamiltonianin}
H_{in} =-\frac{1}{2\epsilon} \sum_{j=0}^{n-1}\left\{-f_j\sigma_j^x \sigma_{j+1}^x+\tilde{g}_j \sigma_j^z\right\},
\end{align}
where
\begin{align}
\label{eq:transversefield2}
    \tilde{g}_j=-f_j-\epsilon r_s/(2x_j^2)-m\sqrt{-f_j}.
\end{align}
Note that $\sqrt{-f_j}$ is real since $f_j<0$ inside the black hole.

The spin model Hamiltonian for the entire region is then
\begin{align}
    H=H_{in}+H_{out}.
\end{align}

Let us take a look at the behavior of this spin model in the neighborhood of the horizon. These are the sites $j=n-1$ and $j=n$ in the summations in Eqs. \eqref{eq:hamiltonianin} and \eqref{eq:hamiltonianout}, respectively, which give the terms
\begin{align}
-\frac{1}{2\epsilon} \left\{-f_{n-1}\sigma_{n-1}^x \sigma_{n}^x+\tilde{g}_{n-1} \sigma_{n-1}^z+g_n \sigma_n^z\right\}
\end{align}

Notice that $f_n=0$ at the horizon, implying that the spins located at $x_n$ and $x_{n+1}$ are completely decoupled. As a result, the Hamiltonian describing the region inside the black hole, $0\leq j\leq n$, becomes dynamically disconnected from the Hamiltonian governing the exterior region. One immediate consequence is that excitations cannot propagate across the horizon in either direction. While the absence of propagation from the interior to the exterior is consistent with the causal structure of a black hole, the converse restriction is unphysical: fields originating outside the black hole should, in principle, be able to cross the horizon and fall inward. This behavior is therefore not faithfully reproduced by the present spin model, which forbids propagation across the horizon altogether. The origin of this limitation can be traced to the coordinate singularity of the Schwarzschild coordinates used in constructing the model. Since the Schwarzschild metric becomes singular at the horizon, the corresponding spin couplings vanish there, artificially severing the connection between the interior and exterior regions. This implies that the present spin model is suitable only for describing fermionic dynamics restricted either to the exterior or to the interior region of the black hole, but not for processes involving propagation across the horizon.

For Majorana fermions outside and far from the black hole, $x\to\infty$, the Hamiltonian Eq. \eqref{eq:hamiltonianout} becomes
\begin{equation}
    H=-\frac{1}{2\epsilon}\sum_{j=1}^{N} {\left\{\sigma_j^x \sigma_{j+1}^x+\left(1 -m\epsilon \right)\sigma_j^z\right\}},
\end{equation}
which is the well-known quantum Ising chain in a transverse field \cite{Sachdev2011QPT}, with an exchange constant $J=1/(2\epsilon)>0$ and a coupling constant $g=1-m\epsilon>0$. We also get the same model if we take the limit as $r_s\to 0$, when the black hole shrinks.

Another disadvantage of using the Schwarzschild coordinates to generate the spin models is the presence of both site-dependent exchange coupling and transverse magnetic field as we have seen in Eqs. \eqref{eq:hamiltonianin} and \eqref{eq:hamiltonianout}. This can complicate the analysis and make future simulations challenging. This motivates us to explore other starting coordinate systems, specifically, the two canonical coordinate systems used in black hole study: the tortoise and Kruskal coordinates. Before we do so, we first explore the spin models from the more general conformally flat coordinates.

\subsection{Spin models of conformally flat metrics}
We will show in this section that we can choose a specific spin model with site-independent spin-exchange coupling from the family of spin models generated from the conformally flat coordinates. We will then exploit this result to calculate the Hawking radiation in Section V.

The conformally flat metrics in two-dimensional spacetime have the form
\begin{equation}
    ds^s=\Omega^2(t,x)(-dt^2+dx^2),
\end{equation}
where $\Omega(t,x)$ is called the conformal factor. Comparing with Eq. \eqref{eq:admmetric}, the metric functions are
\begin{equation}
\label{eq:alphagammaOmega}
    \alpha(t,x)=\gamma(t,x)=\Omega(t,x),\quad
    \beta(t,x)=0.
\end{equation}
Substituting this into Eq. \eqref{nozeta}, we obtain the Hamiltonian
\begin{align}
\label{conformspin}
    H=-\frac{1}{4\epsilon}\sum_{j=1}^{N}
    \Big\{(p+1)\sigma_j^x \sigma_{j+1}^x
    +(p-1)\sigma_j^y \sigma_{j+1}^y\nonumber\\
    +[2p-\epsilon(\partial_x p+2m\Omega)]\sigma_j^z\Big\},
\end{align}
describing a family of spin models parametrized by the free function $p$.

We focus on the particular spin model with $p(t,x)=1$. This removes the $\sigma^y_j\sigma^y_{j+1}$ coupling giving us the Ising Hamiltonian with site-dependent transverse magnetic field:
\begin{align}
    H=-\frac{1}{2\epsilon}\sum_{j=1}^{N}
    \Big\{\sigma_j^x \sigma_{j+1}^x
    +[1-m\epsilon\Omega(t,x)]\sigma_j^z\Big\}. \label{conformal_TImodel}
\end{align}
Alternatively, one can also choose $p(t,x)=-1$ and get rid of the $\sigma^x_j\sigma^x_{j+1}$ interaction. We emphasize that, unlike the Schwarzschild spin models, the conformally flat spin models can be transformed such that the exchange coupling becomes site-independent. This feature is advantageous, as it is generally easier to engineer spatial variations in the magnetic field than to implement position-dependent exchange interactions. 

We further note that the conformal metric considered here is quite general and enables the construction of spin models in a wide range of coordinate systems, including tortoise and Kruskal coordinates for Schwarzschild and Reissner–Nordström black holes. These constructions can be further extended to black holes embedded in de Sitter and anti–de Sitter spacetimes, as well as to cosmological backgrounds such as the FLRW metric studied in Ref.~\cite{Kinoshita2025}. Remarkably, the resulting Majorana quantum field theories in these diverse curved spacetimes can be simulated by a simple spatial engineering of the transverse magnetic field in the transverse Ising model.

For spacetimes with zero cosmological constant, Eq. \eqref{conformal_TImodel} reduces to the standard transverse-field Ising model with constant exchange coupling and magnetic field far from the black hole. This is because the conformal factor approaches unity far from any localized source of curvature, effectively recovering the flat Minkowski spin model. In contrast, de Sitter spacetimes feature an additional, non-local source of curvature: the cosmological horizon. As a result, Eq. \eqref{conformal_TImodel} does not reduce to the ordinary transverse-field Ising model in these cases. A notable example is the Schwarzschild–de Sitter spacetime, which contains both a black hole event horizon and a cosmological horizon, each contributing to the curvature profile and modifying the spin model accordingly.
\subsection{Spin models from tortoise coordinates}
We will now study the spin models generated from various black hole coordinates using the result of the previous subsection. We start with the black hole metric in the tortoise coordinates, which is given by 
\begin{equation}
    ds^2=f(x)[-dt^2+(dx^*)^2],
\end{equation}
where $f(x)=1-r_s/x$ is the usual blackening factor. Note that here $x$ is Schwarzschild spatial coordinate, which is an implicit function of the tortoise coordinate $x^*$ to be given shortly.

Just like in the Schwarzschild coordinates considered above, we divide the regions into inside and outside the black hole. Outside, we perform the transformation
\begin{align}
    x^*=x+r_s\ln\left(\frac{x}{r_s}-1\right),\;\;r_s<x<\infty
\end{align}
where the new spatial coordinate $x^*$ is defined in the interval $-\infty<x^*<\infty$.

In contrast, inside the black hole, we have
\begin{align}
\label{eq:tortoiseschwarzschild}
    x^*=x+r_s\ln\left(1-\frac{x}{r_s}\right),\;\; 0<x<r_s.
\end{align}

In either regions, the black hole horizon is located at $x^*\to -\infty$. This coordinate atlas has a conformally flat metric, with a conformal factor 
\begin{align}
    \Omega(t,x^\ast)=
    \begin{cases}
        \sqrt{f(x)}, & r_s<x<\infty\\
        \sqrt{-f(x)}, & 0<x<r_s,
    \end{cases}
\end{align}
where $x$ is understood to be written as function of the tortoise spatial coordinate $x^\ast$. The tortoise coordinate system also generates a family of spin models, but we are interested in the particular model with $p(t,x)=1$, giving us the Hamiltonian
\begin{align}
\label{tortoisespinmodel}
    H_{out}=-\frac{1}{2\epsilon}\sum_{j=-\infty}^{\infty} {\left\{\sigma_j^x \sigma_{j+1}^x+\left[1 -m\epsilon\sqrt{f[x(x^*_j)]}\right]\sigma_j^z\right\}}.
\end{align}

Note that the blackening factor is written in terms of the tortoise coordinate $x_j^*$, since $x^*$, rather than $x$, is the coordinate being discretized. Accordingly, the spin index $j$ labels lattice sites in the discretized $x^*$ coordinate. Since $x^*$ ranges over the entire real line, $x^*\in(-\infty,\infty)$, the summation over $j$ must also run from $-\infty$ to $\infty$.

The corresponding transverse field is plotted in Fig.~\ref{fig:P1_ret_array}(b). Although the lattice is defined in terms of the tortoise coordinate $x^*$, we plot the transverse field as a function of the Schwarzschild radial coordinate $x$ in order to display its behavior near the horizon, $x\to 1$, in a more compact and physically transparent manner. For a massless Majorana fermion, the transverse field is uniform across the lattice. In contrast, away from the horizon, heavier Majorana fermions require progressively smaller values of the transverse field, which can become negative for sufficiently large masses. As the horizon is approached, corresponding to decreasing $x$, the influence of the spacetime curvature becomes increasingly pronounced and is reflected in the growth of the transverse field. For sufficiently massive Majorana fermions, the transverse field undergoes a sign change, highlighting the interplay between the fermion mass and the gravitational background.

Inside the black hole, we have
\begin{align}
\label{eq:tortoisespinmodelinside}
    H_{in}=-\frac{1}{2\epsilon}\sum_j {\left\{\sigma_j^x \sigma_{j+1}^x+\left[1 -m\epsilon\sqrt{-f(x^*)}\right]\sigma_j^z\right\}}.
\end{align}

This result requires careful interpretation. Inside the black hole, $f(x^*)<0$, and the causal roles of $t$ and $x^*$ are reversed: $t$ becomes a spacelike coordinate, while $x^*$ assumes the role of the timelike coordinate. Consequently, the spin model in Eq.~\eqref{eq:tortoisespinmodelinside} should be understood as a transverse Ising model with a time-dependent transverse field. Moreover, the lattice index $j$ labels discretized values of the coordinate $t$, rather than spatial positions, so that the spin chain effectively represents a discretization of the spacelike direction inside the horizon.

Note that, in both regions, the horizon remains inaccessible, reflecting the fact that it is mapped to $x^*\to -\infty$ in tortoise coordinates. Consequently, the horizon lies at an infinite coordinate distance and cannot be reached within the lattice description. Therefore, much like the spin models constructed from Schwarzschild coordinates, those based on tortoise coordinates are suitable only for simulating Majorana QFTs restricted entirely to either the exterior or the interior of the black hole, but not for processes that involve crossing the horizon.

\subsection{Spin models from Kruskal coordinates}
Another conformal coordinate system for the Schwarzschild black hole that removes the coordinate singularity at the event horizon is the Kruskal coordinate system, in which the metric takes the form:
\begin{align}
    ds^2=k(x)(-d\tilde{t}^2+d\tilde{x}^2),
\end{align}
where $(\tilde{t},\tilde{x})$ are the Kruskal coordinates \cite{MukhanovWinitzki2004}, which are related to the tortoise and Schwarzschild coordinates discussed previously via
\begin{align}
    \tilde{t}=&2r_se^{\frac{x^*}{2r_s}}\sinh\left(\frac{t}{2r_s}\right)\\
    \tilde{x}=&2r_se^{\frac{x^*}{2r_s}}\cosh\left(\frac{t}{2r_s}\right),
\end{align}
and the conformal factor in terms of the Schwarzschild spatial coordinate is given by
\begin{align}
\label{kruskal_factor}
   k(x)=\frac{r_s}{x}e^{1-\frac{x}{r_s}}.
\end{align}
The Kruskal coordinate system corresponds to a freely falling observer and, unlike the previously discussed coordinate systems, extends smoothly across the event horizon to cover the black hole’s interior. This feature is also reflected in the corresponding spin model, as we will now demonstrate.

The conformal factor of the Kruskal metric is $\Omega(t,x)=\sqrt{k(x)}$ and from Eq. \eqref{conformal_TImodel} the Hamiltonian of the Kruskal spin model is
\begin{align}
\label{kruskal}
    H=-\frac{1}{2\epsilon}\sum_{j=1}^{L} {\left\{\sigma_j^x \sigma_{j+1}^x+\left[1 -m\epsilon\sqrt{k_j}\right]\sigma_j^z\right\}},
\end{align}
where $k_j=k[x(\tilde{t},\tilde{x}_j)]$ is the discretized version of Eq. \eqref{kruskal_factor}. Note that implicitly, the spatial Schwarzschild coordinate $x$ also depends on the Kruskal time, but only the Kruskal space $\tilde{x}_j$ is discretized.

We present in Fig.~\ref{fig:P1_ret_array}(c) the transverse magnetic field as a function of the Schwarzschild spatial coordinate, for ease of comparison with the other spin models. As in the tortoise-coordinate construction, a massless Majorana fermion in the black-hole background is reproduced by a uniform transverse field. The differences between the curves become increasingly pronounced as the horizon is approached, reflecting the growing influence of spacetime curvature in this region.

Far from the black hole, the conformal factor $\Omega(t,x)$ tends to zero. As a result, the transverse magnetic field
\begin{align}
\label{eq:transversekruskal}
g_K(x)=1-m\epsilon\sqrt{k(x)},
\end{align}
approaches unity, which corresponds to the quantum critical point of the spin model. In the continuum limit, the Kruskal spin model in this asymptotic region thus reduces to the quantum field theory of an effectively massless Majorana fermion, since the mass term vanishes in the Hamiltonian density. This can be seen explicitly in Eq.~\eqref{eq:QFT}, where the mass contribution $m\alpha$ contains the metric function $\alpha(x)=\sqrt{k(x)}$, originating from the factor $\sqrt{-g}$ in the Hamiltonian and Lagrangian densities in Eqs.~\eqref{eq:lagrangian1} and \eqref{eq:QFT}, respectively. Because this factor decays asymptotically, the mass term is effectively suppressed at large distances from the black hole. This behavior is illustrated in Fig.~\ref{fig:P1_ret_array}(c), where all curves approach the asymptotic value $g_K(x)\to 1$ as $x$ increases. This is consistent with the well-known result that the continuum limit of the quantum-critical Ising model is described by a quantum field theory of massless Majorana fermions in flat spacetime \cite{Sachdev2011QPT}.

Before proceeding, we comment further on the transverse fields in the tortoise and Kruskal models. The transverse field is determined by three parameters: the fermion mass $m$, the lattice spacing $\epsilon$, and the spacetime curvature through the factors $\sqrt{k}$ or $\sqrt{f}$. In this analogue spin system, the lattice spacing plays the role of an effective \lq\lq Planck length\rq\rq. If the spin system is regarded as a low-energy analogue of the Standard Model fermions, then the relevant regime is $m\epsilon\ll 1$. As shown in Figs.~\ref{fig:P1_ret_array}(b) and (c), the transverse field remains strictly positive in this regime. In contrast, as the fermion mass approaches and exceeds the effective Planck scale, $m\epsilon\gtrsim 1$, the transverse field undergoes a sign reversal.

\section{Emergent general covariance}
\label{sec:diffeomorphism}
In the previous section, we demonstrated that distinct spin models arise from different choices of coordinate systems. This naturally leads to the question: Do these different spin models describe the same Majorana field theory in a black hole background in the continuum limit? The principle that the physical content of a theory remains invariant under arbitrary coordinate transformations is known as general covariance, or diffeomorphism invariance, and constitutes one of the foundational principles of Einstein's general theory of relativity (for a broader discussion of its conceptual status, see Refs.~\cite{Sorkin2002Kretschmann,Norton2003GeneralCovariance}).

We now show that general covariance emerges in the continuum limit of the distinct microscopic spin models constructed using different coordinate systems. This is evidenced by the invariance of the effective action under coordinate transformations.

\subsection{The dictionary as a covariant map}
The foundation of this argument is the dictionary established in Section II between the spin model parameters and the metric functions of the curved spacetime. For a generic two-dimensional curved spacetime with ADM metric Eq.~\eqref{eq:admmetric}, the Majorana QFT Hamiltonian density Eq.~\eqref{eq:QFT}, or equivalently Eq.~\eqref{continuum_nozeta} with $\zeta=0$, depends on the metric only through the three combinations:
\begin{align}
\label{eq:metriccombinations}
    \frac{\alpha(t,x)}{\gamma(t,x)}, \qquad \beta(t,x),\qquad m\alpha(t,x).
\end{align}
These are the only metric-dependent quantities that appear in the continuum Hamiltonian density. The spin model parameters $J^{ab}_j$ and the transverse field $h_j$ are determined by these three combinations together with the free function $p(t,x)$, which encodes a microscopic non-uniqueness that disappears in the continuum limit.

The dictionary is derived under the assumption that the spacetime metric admits an ADM decomposition with a spacelike foliation, as in Eq.~\eqref{eq:admmetric}. This is the natural domain of the Kinoshita et al. construction~\cite{Kinoshita2025}, and it is the reason we restrict \lq\lq coordinate transformation\rq\rq throughout this section to mean any smooth diffeomorphism between two ADM-compatible coordinate systems covering the same spacetime patch. Transformations that exchange the causal roles of coordinates across a horizon, as occurs in the passage between Schwarzschild and Kruskal coordinates at $x=r_s$, fall outside this scope and must be handled by constructing separate spin models in each ADM-foliable patch, as done in Section III.




\begin{figure*}[t]
    \centering

    \includegraphics[trim=0.16cm 5.69cm 0cm 6.35cm, clip, width=1.026\textwidth]{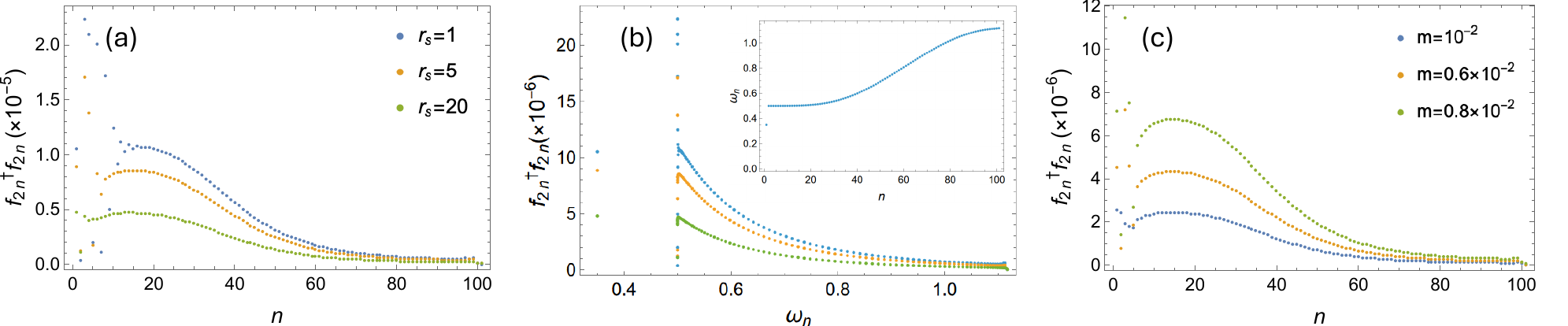}

    \caption{Numerical simulation of the fermion number of the Kruskal vacuum using the Ising model, as measured by an observer in tortoise coordinates. Panels (a) and (b) show the fermion number for varying Schwarzschild radius $r_s$ as a function of the mode number and the energy eigenvalue, respectively. The inset in (b) shows the energy eigenvalues as a function of the mode number. Panel (c) shows the fermion number as a function of the mode number for varying Majorana mass $m$.
}
    \label{fig:Particleproduction}
\end{figure*}

\subsection{Two sources of microscopic non-uniqueness}\label{subsec:two_sources}
There are two independent sources of non-uniqueness in the spin model construction. The first is coordinate freedom, which means that a given spacetime admits infinitely many coordinate systems. Each choice of ADM-compatible coordinates yields a distinct set of metric functions ($\alpha,\beta,\gamma$) and, through the dictionary, a distinct spin Hamiltonian. The Schwarzschild, tortoise, and Kruskal Hamiltonians of the previous section demonstrate this directly.

The second is what we will call the $p$-freedom. This means that even for a particular choice of coordinate system, the function $p(t,x)$ can be chosen arbitrarily without affecting the continuum theory. This generates a further family of distinct microscopic spin Hamiltonians, all flowing to the same Majorana QFT in the continuum limit. In the spin model language, $p(t,x)$ parametrizes distinct UV completions of the same low-energy physics; it has no geometric counterpart in the continuum description.

Both sources of non-uniqueness leave the continuum theory invariant. We now demonstrate this explicitly for coordinate freedom, which is more relevant in studying QFTs in curved spacetimes.

\subsection{Diffeomorphism invariance in the continuum theory}
Consider a general coordinate transformation $(t,x)\to(\tilde{t},\tilde{x})$ between two ADM-compatible coordinate systems covering the same spacetime patch. The metric tensor transforms as
\begin{align}
\label{eq:metric_transformation}
    \tilde{g}_{\mu\nu}=\frac{\partial x^\alpha}{\partial x^\mu}\frac{\partial x^\beta}{\partial x^\nu}g_{\alpha\beta},
\end{align}
from which the new metric functions $(\tilde{\alpha},\tilde{\beta},\tilde{\gamma})$ are read off via
\begin{align}
\label{new_metric_func}
    \tilde{\alpha}=\sqrt{\frac{\tilde{g}_{\tilde{t}\tilde{x}}^2}{\tilde{g}_{\tilde{x}\tilde{x}}}-\tilde{g}_{\tilde{t}\tilde{t}}},\quad
    \tilde{\beta}=-\frac{\tilde{g}_{\tilde{t}\tilde{x}}}{\tilde{g}_{\tilde{x}\tilde{x}}},\quad
    \tilde{\gamma}=\sqrt{\tilde{g}_{\tilde{x}\tilde{x}}},
\end{align}
all of which are functions of the new coordinates.

The Majorana field transforms as
\begin{align}
\label{eq:Majorana_field_transformation}
    \tilde{\Psi}(\tilde{t},\tilde{x})=\left|\frac{\partial x}{\partial \tilde{x}}\right|^{1/2}\Psi(t,x),
\end{align}
which is the unique transformation rule that preserves the canonical anticommutation relations of the Majorana field in both coordinate systems.

We now show that the action
\begin{align}
    S = \int dt\,dx\,\mathcal{L}(t,x)
\end{align}
is invariant under this transformation, i.e., $S = \tilde{S}$. Here, the Lagrangian density is given by
\begin{align}
\label{eq:Lagrangian}
    \mathcal{L}=
    &\frac{i}{2} (\Psi^{\dagger} \partial_t \Psi+\Psi \partial_t \Psi^\dagger)
    +\frac{\alpha}{2\gamma} (\Psi^{\dagger} \partial_x \Psi^{\dagger}-\Psi \partial_x \Psi)\nonumber\\
    &+i\frac{\beta}{2} (\Psi^{\dagger} \partial_x \Psi+\Psi \partial_x \Psi^{\dagger})
    +m\alpha\Psi^{\dagger} \Psi.
\end{align}
The key observation is that the Lagrangian density $\mathcal{L}$ in Eq.~\eqref{eq:Lagrangian} is not a scalar: it is obtained from
the covariant Lagrangian $\mathcal{L}_{\mathrm{cov}}$ by absorbing the
factor $\sqrt{-g} = \alpha\gamma$,
\begin{equation}
    \mathcal{L}(t,x) = \sqrt{-g}\,\mathcal{L}_{\mathrm{cov}}.
\end{equation}
Under the coordinate transformation, $\sqrt{-g} = \alpha\gamma$
transforms as $\sqrt{-\tilde{g}} = \tilde{\alpha}\tilde{\gamma}$, while
$\mathcal{L}_{\mathrm{cov}}$ is a true scalar density of weight zero.
Consequently, $\mathcal{L}$ transforms as a scalar density of weight one:
\begin{align}
    \mathcal{L}(t,x) = \left|\frac{\partial \tilde{x}}{\partial x}\right|
    \tilde{\mathcal{L}}(\tilde{t},\tilde{x}).
    \label{eq:Ldensity}
\end{align}
The volume element transforms in the opposite sense,
\begin{align}
    dt\,dx = \left|\frac{\partial \tilde{x}}{\partial x}\right|^{-1}
    d\tilde{t}\,d\tilde{x}.
    \label{eq:volumetransform}
\end{align}
Substituting both into the action,
\begin{align}
    S &= \int dt\,dx\,\mathcal{L}(t,x)\nonumber\\
    &= \int \left|\frac{\partial\tilde{x}}{\partial x}\right|^{-1}
    d\tilde{t}\,d\tilde{x}
    \cdot \left|\frac{\partial\tilde{x}}{\partial x}\right|
    \tilde{\mathcal{L}}(\tilde{t},\tilde{x})\nonumber\\
    &= \int d\tilde{t}\,d\tilde{x}\,\tilde{\mathcal{L}}(\tilde{t},\tilde{x})\nonumber\\
    &= \tilde{S}.
\end{align}
That is, the Jacobian factors cancel exactly, leaving $S = \tilde{S}$.

To verify Eq.~\eqref{eq:Ldensity} directly from the form of $\mathcal{L}$ in Eq.~\eqref{eq:Lagrangian}, consider for concreteness
the purely spatially reparametrization $\tilde{t} = t$, $\tilde{x} =
\tilde{x}(x)$, with $J \equiv d\tilde{x}/dx$. One finds $\tilde{\alpha}
= \alpha$, $\tilde{\beta} = J\beta$, $\tilde{\gamma} = \gamma/J$, and
the field rescaling Eq.~(\ref{eq:Majorana_field_transformation}) gives $\Psi =
J^{1/2}\tilde{\Psi}$. Each bilinear in $\mathcal{L}$ then picks up a
factor $J$ from the two field rescalings ($J^{1/2}$ each), and
the spatial derivative $\partial_x = J\,\partial_{\tilde{x}}$ brings a
factor $J$ while the modified coefficient $\tilde{\alpha}/\tilde{\gamma}
= J(\alpha/\gamma)$ brings another $J$, yielding a net factor of $J$ per
term, as required by Eq.~(\ref{eq:Ldensity}).

In the $\alpha/\gamma$ term, the gradient terms of the form
$(\partial_{\tilde{x}} J^{1/2})\Psi\Psi$ and $(\partial_{\tilde{x}} J^{1/2})\Psi^\dagger\Psi^\dagger$ arising from
differentiating the position-dependent rescaling vanish identically
by Grassmann anticommutativity: $\Psi \Psi = 0$ and $\Psi^\dag \Psi^\dag = 0$, respectively. In the $\beta$ term, the bilinear $\Psi^{\dagger} \partial_x \Psi+\Psi \partial_x \Psi^{\dagger}$ gains two extra terms: $(\partial_{\tilde{x}} J^{1/2})(\Psi^\dagger\Psi+\Psi\Psi^\dagger)$, whose sum vanishes due to the Grassmann anticommutativity: $\Psi^\dag\Psi=-\Psi\Psi^\dag$. These make
the transformation Eq.~(\ref{eq:Majorana_field_transformation}) exact.

As a concrete check of the entire argument, consider the exterior
Schwarzschild-to-tortoise transition. The tortoise coordinate
$x^* = x + r_s \ln(x/r_s - 1)$ has Jacobian $J = dx^*/dx = f(x)^{-1}$
where $f(x) = 1 - r_s/x$. From the Schwarzschild ADM functions
$\alpha = \sqrt{f}$, $\beta = 0$, $\gamma = 1/\sqrt{f}$ one obtains,
via Eqs.~(\ref{new_metric_func}),
\begin{equation}
    \tilde{\alpha} = \sqrt{f},\quad
    \tilde{\beta} = 0,\quad
    \tilde{\gamma} = \frac{1\sqrt{f}}{1/f} = \sqrt{f},
\end{equation}
which is precisely the conformally flat structure $\tilde{\alpha} =
\tilde{\gamma} = \Omega(x^*)$ with $\Omega = \sqrt{f}$, as used in
Section~III\,C. The three combinations entering
Eq.~(\ref{eq:metriccombinations}) transform as $\alpha/\gamma = f \to
\tilde{\alpha}/\tilde{\gamma} = 1$ and $m\alpha = m\sqrt{f} \to
m\tilde{\alpha} = m\sqrt{f}$ (unchanged), confirming that the continuum
Hamiltonian density in tortoise coordinates is a coordinate-transformed
version of the Schwarzschild one, not a new field theory.

\subsection{Emergence of diffeomorphism invariance}
The picture that emerges is the following. At the microscopic level, each spin Hamiltonian in our family is a non-relativistic lattice system with no spacetime symmetry in general. The Hamiltonian is simply a sum of local spin-exchange couplings and on-site transverse fields with prescribed spatial dependence. The lattice sites are labeled by integers; the parameters $J^{ab}_j$ and $h_j$ are fixed at each site; and there is no microscopic notion of coordinate transformation.

In the continuum limit $\epsilon\to 0$, however, two microscopically distinct spin Hamiltonians, say, one constructed from Schwarzschild coordinates and another from tortoise coordinates in the exterior of the event horizon, flow to continuum Lagrangians that are related by a coordinate transformation. By the coordinate covariance of the Majorana action proved above, these two continuum theories describe the same physical Majorana QFT on the \textit{same} physical black hole spacetime. The diffeomorphism invariance manifest in the continuum was absent at the microscopic level: it emerges dynamically as a property of the low-energy description.

This is a concrete instance of a more general phenomenon: spacetime symmetries emerge from microscopic theories that do not possess them. Emergent Lorentz invariance \cite{Kharuk2016} and emergent gauge symmetry \cite{wen2004quantum, emergentgauge} are familiar phenomena in quantum many-body systems. Emergent general covariance, the full diffeomorphism group of a curved spacetime, is considerably rarer. Our construction provides an explicit realization: the microscopic spin models are non-relativistic and possess no geometric structure, yet their continuum limit reproduces a fully covariant QFT in curved spacetime.

This perspective also clarifies the conceptual status of the two non-uniquenesses identified Section~\ref{subsec:two_sources}. Coordinate freedom corresponds to a geometric reparametrization of the same physical spacetime, with a clear continuum interpretation as the action of the diffeomorphism group. The $p$-freedom, in contrast, is a purely microscopic redundancy: it has no geometric counterpart in the continuum theory, and parametrizes distinct ultraviolet completions of the same low-energy Majorana QFT. Together, the two freedoms delineate the full space of microscopic spin models compatible with a given Majorana QFT in a given curved spacetime.

\section{Simulating Black hole Particle production}
One of the most striking consequences of QFT in a black hole background is particle production, which leads to Hawking radiation and eventual black hole evaporation. However, these effects are exceedingly weak and remain inaccessible to direct experimental observation. In this section, we show how black hole particle production can be simulated using the spin models derived in this work.

The key idea underlying particle production is the inequivalence of vacuum states. Let $\{d_a,d_a^\dagger\}$ and $\{f_a,f_a^\dagger\}$ denote the operator bases that diagonalize the Kruskal and tortoise Hamiltonian spin models, respectively:
\begin{align}
H_K &= \sum_a \varepsilon^K_a d_a^\dagger d_a + E^K_0\\
\label{eq:diagonaltortoise}
H_T &= \sum_a \varepsilon^T_a f_a^\dagger f_a + E^T_0.
\end{align}
The corresponding vacuum states are defined by
\begin{align}
\label{eq:vacuum}
d_a|\Omega_K\rangle = 0,\qquad
f_a|\Omega_T\rangle = 0.
\end{align}

The notion of vacuum is observer-dependent: a state that appears as a vacuum in one frame is, in general, not empty when described in another. More concretely, consider the number operator in tortoise coordinates,
\begin{align}
\label{eq:particlenumberkruskalvac}
\hat{n}^T_a \equiv f_a^\dagger f_a,
\end{align}
and evaluate its expectation value in the Kruskal vacuum $|\Omega_K\rangle$. We find
\begin{align}
\label{eq:particlenumberkruskalvac2}
\langle \Omega_K | \hat{n}^T_a | \Omega_K \rangle
\equiv \langle \Omega_K | f_a^\dagger f_a | \Omega_K \rangle \neq 0.
\end{align}
Thus, an observer natural to the tortoise coordinates will perceive the Kruskal vacuum as containing particle excitations.

We now discuss how this phenomenon can be simulated. We begin with an Ising system whose spatially varying transverse field is engineered such that it is described by the Kruskal Hamiltonian in Eq.~\eqref{kruskal}. The system is then prepared in its ground state, which corresponds to the vacuum of its fermionic excitations. This state is precisely the Kruskal vacuum, $|\Omega_K\rangle$.

The Kruskal Hamiltonian differs from the tortoise spin model in Eq. \eqref{tortoisespinmodel} only through the transverse field. We therefore perform a quench, say at $t=0$, of the transverse field from the Kruskal form in Eq. \eqref{eq:transversekruskal} to the tortoise form,
\begin{align}
g_T(x^*) = 1 - m\epsilon \sqrt{f(x^*)}.
\end{align}

Following this quench, the system now evolves under the tortoise spin Hamiltonian for $t>0$. In tortoise coordinates, the state $|\Omega\rangle_K$ is not the ground state and contains particle excitations, as quantified by the expectation value in Eq. \eqref{eq:particlenumberkruskalvac2}. Note that the number operator $f^\dagger_af_a$ is stationary for $t>0$ since it is evolved (in the Heisenberg picture) by Eq. \eqref{eq:diagonaltortoise}, so that Eq. \eqref{eq:particlenumberkruskalvac2} is time independent. We now relate this particle number to observables of the underlying spin model. Using the Jordan–Wigner transformation in Eq. \eqref{eq:jordanwigner}, the spin Hamiltonians can be written in fermionic form as
\begin{align}
\label{eq:spinmodelsfermionoperators}
H = -\frac{1}{2}\sum_{i,j}
(c_i^\dagger, c_i)
\begin{pmatrix}
A_{ij} & \Delta_{ij}\\
(\Delta^T)_{ij} & -A_{ij}
\end{pmatrix}
\begin{pmatrix}
c_j\\
c_j^\dagger
\end{pmatrix}.
\end{align}

Here, the explicit matrix elements of the blocks are
\begin{align}
A_{ij}=&\frac{1}{2}(\delta_{i,j+1}+\delta_{i,j-1})-\delta_{i,j}g_i\\
\Delta_{ij}=&\delta_{i,j-1},
\end{align}
where $g_j$ is the corresponding transverse field for either the tortoise or Kruskal coordinates.

The Hamiltonian can be diagonalized via a unitary transformation, corresponding to a generalized Bogoliubov transformation. For the tortoise spin model, this transformation takes the form
\begin{align}
\label{eq:bogoliubovtransform}
\begin{pmatrix}
f_{1a}\\
f_{2a}^\dagger
\end{pmatrix}
=\sum_j
\begin{pmatrix}
[T_{11}]_{aj} & [T_{12}]_{aj}\\
[T_{21}]_{aj} & [T_{22}]_{aj}
\end{pmatrix}
\begin{pmatrix}
c_j\\
c_j^\dagger
\end{pmatrix},
\end{align}
which brings the Hamiltonian into diagonal form as in Eq. \eqref{eq:diagonaltortoise}.

Before calculating the particle number, we first discuss a subtle aspect of the Bogoliubov transformation in Eq. \eqref{eq:bogoliubovtransform}. The transformed theory contains two types of quasiparticles described by the creation operators, $f^\dagger_{1j}$ and $f^\dagger_{2j}$, distinguished by the additional indices 1 and 2 besides the lattice index $j$. At first sight, this appears to double the number of degrees of freedom. This doubling, however, is an artifact of enlarging the Hilbert space to the particle-hole (Nambu) space, as is standard in the theory of superconductivity. As we show below, the operators $f^\dagger_{1j}$ and $f^\dagger_{2j}$ satisfy a constraint that removes the redundant degrees of freedom, thereby restricting the enlarged Hilbert space to its physical subspace. To derive this constraint we invert Eq. \eqref{eq:bogoliubovtransform}, which we can write in the form
\begin{align}
\label{eq:inversebogol1}
    c_j=&[T^{-1}_{11}]_{ja}f_{1a}+[T^{-1}_{12}]_{ja}f_{2a}^\dagger\\
\label{eq:inversebogol2}
    c_j^\dagger=&[T^{-1}_{21}]_{ja}f_{1a}+[T^{-1}_{22}]_{ja}f_{2a}^\dagger.
\end{align}
Here, we also write the inverse of the $T$ matrix in Eq. \eqref{eq:bogoliubovtransform} into particle-hole blocks $T^{-1}_{11}$, $T^{-1}_{12}$, etc. For each block, $[T^{-1}]_{ja}$ means the $j$th row and $a$th column of that matrix block.

The Hermitian conjugate of Eq. \eqref{eq:inversebogol1} must coincide with Eq. \eqref{eq:inversebogol2}. Enforcing this condition yields
\begin{align}
\label{eq:operatorrelation}
[T^{-1}_{11}]_{ja}f^\dagger_{1a}+[T^{-1}_{12}]_{ja}f_{2a}=[T^{-1}_{21}]_{ja}f_{1a}+[T^{-1}_{22}]_{ja}f^\dagger_{2a}.
\end{align}
Equation \eqref{eq:operatorrelation} shows that the operators $f^\dagger_{1a}$ and $f^\dagger_{2a}$ are not independent, but are constrained by the above relation. Equivalently, the physical Hilbert space is identified as the subspace of the enlarged Hilbert space whose states satisfy the operator identity Eq. \eqref{eq:operatorrelation}. Consequently, it is sufficient to focus on only one type of quasiparticle, which we take to be $f^\dagger_{2a}$. The number operator can then be written as
\begin{align}
    f^\dagger_{2a}f_{2a}
    =&\sum_{i,j}\bigg([T_{21}]_{ai}c_i +[T_{22}]_{ai}c^\dagger_i\bigg)\nonumber\\
    &\times\bigg([T_{22}]_{aj}c_j +[T_{21}]_{aj}c^\dagger_j\bigg).
\end{align}

Using the inverse of the Jordan-Wigner transformation,
\begin{align}
    c_j=\bigg(\prod_{k=1}^{j-1}\sigma^z_k\bigg)\sigma^+_j,\;\;\;c^\dagger_j=\bigg(\prod_{k=1}^{j-1}\sigma^z_k\bigg)\sigma^-_j,
\end{align}
and taking the expectation value of $\hat{n}^T_a$ using the state $|\Omega_K\rangle$, which we will denote by $\langle\cdot\cdot\cdot\rangle_K$ for notational brevity, we have
\begin{align}
\label{eq:particleproductionspin}
    n_a=&\sum_{a,i,j}\bigg([T_{21}]_{ai}[T_{22}]_{aj}\left\langle P_{ij}\sigma^+_i\sigma^+_j\right\rangle_K\nonumber\\
    &+[T_{21}]_{ai}[T_{21}]_{aj}\left\langle P_{ij}\sigma^+_i\sigma^-_j\right\rangle_K\nonumber\\
    &+[T_{22}]_{ai}[T_{22}]_{aj}\left\langle P_{ij}\sigma^-_i\sigma^+_j\right\rangle_K\nonumber\\
    &+[T_{22}]_{ai}[T_{21}  ]_{aj}\left\langle P_{ij}\sigma^-_i\sigma^-_j\right\rangle_K\bigg),
\end{align}
where
\begin{align}
    P_{ij}\equiv\prod_{n=1}^{i-1}\prod_{m=1}^{j-1}\sigma^z_n\sigma^z_m.
\end{align}
The spin correlations $\langle P_{ij}\sigma^\alpha_i\sigma^\beta_j\rangle_K$, with $\alpha,\beta\in\{+,-\}$, can be further decomposed in terms of the Hermitian Pauli string operators $\langle P_{ij}\sigma^a_i\sigma^b_j\rangle_K$, with $a,b\in\{x,y\}$, which in turn can be simplified by exploiting the properties of the Pauli operators. This gives us,
\begin{align}
\label{eq:pij}
    P_{ij}=\begin{cases}
        I, &  i=j\\
        \prod_{m=\text{min}(i,j)}^{\text{max}(i,j)-1}\sigma^z_m, & i\neq j.
    \end{cases}
\end{align}
The submatrices $[T]$ are obtained numerically by diagonalizing the $2N\times 2N$ BdG matrix of the tortoise Hamiltonian Eq.~\eqref{tortoisespinmodel}, and are therefore purely classical quantities requiring no quantum resources to compute. Equation \eqref{eq:particleproductionspin}, along with Eq. \eqref{eq:pij}, shows that the fermion production can be simulated by measuring the appropriate spin correlations of the Ising model.

For comparison, let us calculate this particle number numerically. Just as in Eq. \eqref{eq:bogoliubovtransform}, we can diagonalize the Kruskal Hamiltonian via the Bogoliubov transformation
\begin{align}
\label{eq:bogoliubovtransformkruskal}
\begin{pmatrix}
d_{1a}\\
d_{2a}^\dagger
\end{pmatrix}
=\sum_j
\begin{pmatrix}
[K_{11}]_{aj} & [K_{12}]_{aj}\\
[K_{21}]_{aj} & [K_{22}]_{aj}
\end{pmatrix}
\begin{pmatrix}
c_j\\
c_j^\dagger
\end{pmatrix}.
\end{align}

Inverting this transformation, then substituting the result into Eq. \eqref{eq:bogoliubovtransform}, we get
\begin{align}
\label{eq:bogoliubovtortoisekruskal}
\begin{pmatrix}
f_{1n}\\
f_{2n}^\dagger
\end{pmatrix}
=\sum_a
\begin{pmatrix}
[M_{11}]_{na} & [M_{12}]_{na}\\
[M_{21}]_{na} & [M_{22}]_{na}
\end{pmatrix}
\begin{pmatrix}
d_{1a}\\
d_{2a}^\dagger
\end{pmatrix},
\end{align}
where
\begin{align}
\label{eq:matrixm}
\begin{pmatrix}
[M_{11}]_{na} & [M_{12}]_{na}\\
[M_{21}]_{na} & [M_{22}]_{na}
\end{pmatrix}\equiv&\sum_j
\begin{pmatrix}
[T_{11}]_{nj} & [T_{12}]_{nj}\\
[T_{21}]_{nj} & [T_{22}]_{nj}
\end{pmatrix}\nonumber\\
&\times
\begin{pmatrix}
[K_{11}]_{aj} & [K_{12}]_{aj}\\
[K_{21}]_{aj} & [K_{22}]_{aj}
\end{pmatrix}^{-1}.
\end{align}
Equation.~\eqref{eq:bogoliubovtortoisekruskal} gives the transformation from the basis operators $\{d,d^\dagger\}$ to $\{f,f^\dagger\}$, which diagonalize the Kruskal and tortoise Hamiltonians, respectively.

We can now use Eq.~\eqref{eq:bogoliubovtortoisekruskal} and Eq.~\eqref{eq:vacuum} to calculate the expectation value of the tortoise number operator in the Kruskal vacuum. We get the particle number
\begin{align}
\label{eq:particlenumber}
    _K\langle\Omega |f^\dagger_{2n} f_{2n} |\Omega\rangle_K
    =\sum_a\big([M_{21}]_{na}\big)^2,
\end{align}
of the $n$th mode. We can also calculate the number $f^\dagger_{1,n} f_{1,n}$ instead, in which case, we get a hole-like distribution.

We evaluate the right-hand side of Eq.~\eqref{eq:particlenumber} numerically. Figure~\ref{fig:Particleproduction}(a) shows the fermion number as a function of the mode index $n$, where $n=0$ corresponds to the lowest-energy eigenmode, for several values of the Schwarzschild radius $r_s$. Throughout this calculation, we use 101 lattice sites with lattice spacing $\epsilon=1$, Majorana mass $m=10^{-2}$, and Kruskal time $\tilde{t}_K=0.5$. The oscillations observed in the lowest modes are likely attributable to two effects. First, the particle number is evaluated at the finite Kruskal time $\tilde{t}_K=0.5$ following the quench, before the system has fully relaxed to its steady state. Second, finite-size and lattice discretization effects may also contribute to these low-energy oscillations. Nonetheless, with the exception of the lowest-energy modes, the particle number follows the expected profile of Fermi-Dirac distribution. 

Figure~\ref{fig:Particleproduction}(b) shows the fermion number as a function of the energy eigenvalue. Except for the lowest-energy modes, the particle distribution exhibits the exponential decay characteristic of a Fermi-Dirac distribution at finite temperature. Fitting the numerical results to the Fermi-Dirac distribution yields Hawking temperatures of $k_BT/J\sim 0.192$, $0.191$, and $0.190$ for Schwarzschild radii $r_s=1$, $5$, and $20$, respectively, where $k_B$ is the Boltzmann constant and $J$ is the Ising exchange coupling. Thus, the analogue model correctly reproduces the expected trend that the Hawking temperature decreases as the Schwarzschild radius increases. The inset shows the energy eigenvalues as a function of the mode number, revealing that the lowest-energy modes are nearly degenerate.

Figure~\ref{fig:Particleproduction}(c) shows the particle number for different values of the Majorana mass. In this case, we fix $r_s=10$, while keeping all other parameters identical to those used in Fig.~\ref{fig:Particleproduction}(a). We find that the particle number increases with increasing Majorana mass. This behavior can be understood by recalling that, in the lattice Ising model, the effect of spacetime curvature enters through the mass term. Consequently, a larger value of $m$ enhances the influence of the background curvature, leading to greater particle production. In the limiting case $m=0$, the curvature-dependent contribution vanishes, so that the Kruskal and tortoise vacua become identical. We have verified numerically that the particle number vanishes identically in this limit.

To summarize, the simulation of black hole particle production proceeds as follows. We begin with the Kruskal Hamiltonian and prepare its ground state $|\Omega_K\rangle$. We then perform a quench in the transverse field, thereby changing the system into the tortoise spin model. By measuring the appropriate spin correlations that appear in Eq. \eqref{eq:particleproductionspin}, the resulting particle production can be experimentally accessed.
\section{Conclusions and outlook}
We have constructed a family of microscopic spin models that simulate Majorana QFT in curved spacetime using different coordinate choices for the Schwarzschild black hole background. Although these models are distinct at the microscopic level, we show that their continuum limits converge to the same Majorana QFT in a black hole background, up to coordinate transformations. This provides a rare example of general covariance emerging from a condensed matter system. It also suggests the possibility of classifying spin models into a new universality class characterized by emergent general covariance in the continuum limit. In addition, we propose a scheme to simulate particle production in a black hole using spin systems and show that it can be detected through measurements of appropriate spin correlations.

We end this section by discussing the potential experimental platforms for realizing or quantum-simulating these models. A natural solid-state realization is provided by CoNb$_2$O$_6$, which has been experimentally demonstrated to realize the one-dimensional transverse-field Ising model \cite{Coldea2010E8, Ringler2022CoNb2O6}. In this setting, the transverse field corresponds to a physical magnetic field, which may in principle be engineered to acquire the spatial dependence required for simulating Majorana QFT in a black hole background. A key limitation of this platform is that the spin couplings are fixed by the material’s microscopic chemistry and are therefore not readily tunable.

We next consider platforms with tunable spin interactions, in contrast to the fixed-coupling magnetic system CoNb$_2$O$_6$. A prominent example is trapped-ion chains \cite{Blatt2012QuantumSimulations, Britton2012EngineeredIsing, Zhang2017DynamicalPhaseTransition}, where effective spin–spin couplings can be controlled via laser detuning and intensity. In addition, the transverse field is synthetic and can be engineered with spatial profiles using resonant microwave or Raman driving.

Rydberg atom arrays provide another versatile platform \cite{Labuhn2016TunableRydbergArrays, Browaeys2020ManyBodyRydberg, Borish2020TFIMRydbergDressed}. In these systems, spin interactions can be tuned through interatomic spacing, choice of Rydberg state, laser detuning, and dressing strength. The effective transverse field is likewise controllable via laser Rabi frequencies and microwave driving, enabling spatially varying field configurations.

Finally, superconducting qubit arrays constitute a complementary platform for quantum simulation \cite{Georgescu2014QuantumSimulation}. In these architectures, spin interactions can be engineered through tunable coupler circuits, while site-dependent transverse fields can be implemented via individually addressable microwave drives.

Taken together, these platforms demonstrate that although direct observation of QFT phenomena in curved spacetime remains challenging, a variety of highly controllable quantum simulators offer promising routes toward their experimental realization.



\bibliography{apssamp}

\appendix
\clearpage

\end{document}